\documentclass[12pt]{iopart}
\usepackage{bbm}
\usepackage{iopams}  
\input epsf
\eqnobysec

\newcommand{\Vol}  {\mathrm{Vol}}
\newcommand{\Kc} {\mathcal{K}}
\newcommand{\Nc} {\mathcal{N}}
\newcommand{\Oc} {\mathcal{O}}
\newcommand{\Vc} {\mathcal{V}}
\newcommand{\Yc} {\mathcal{Y}}
\newcommand{\id}{\mathbbm{1}}
\newcommand{\Zset}{\mathbbm{Z}}

\newcommand{\Rset}{\mathbbm{R}}

\newcommand{\TilI}{{\tilde{I}}}
\newcommand{\TilJ}{{\tilde{J}}}
\newcommand{\TilK}{{\tilde{K}}}

\newcommand{\TilM}{{\tilde{M}}}
\newcommand{\refeq}[1]{(\ref{#1})}
\def\der{\partial}

\newcommand{\CMP}{{\it Commun.\ Math.\ Phys.\ }}

\usepackage[hypertex]{hyperref}
\begin{document}


\title{The K\"ahler Cone as Cosmic Censor
}
 
\author{Christoph Mayer and  Thomas Mohaupt
}
\address{Theoretisch-Physikalisches Institut,
  Friedrich-Schiller-Universit\"{a}t
  Jena, Max-Wien-Platz 1, D-07743 Jena, Germany
}

\eads{\mailto{C.Mayer@tpi.uni-jena.de}, \mailto{T.Mohaupt@tpi.uni-jena.de}}
   \begin{abstract}
     M-theory effects prevent five-dimensional domain-wall and black-hole
     solutions from developing curvature singularities. While so far this
     analysis was performed for particular models, we now present a
     model-independent proof that these solutions do not have naked
     singularities as long as the K\"ahler moduli take values inside the
     extended K\"ahler cone.  As a by-product we obtain information on the
     regularity of the K\"ahler-cone metric at boundaries of the K\"ahler cone
     and derive relations between the geometry of moduli space and space-time.
 \end{abstract}


\submitto{\CQG}
\pacs{04.20.Dw, 11.25.Yb, 04.65.+e, 11.27.+d, 04.70-s}
\maketitle
\section{Introduction}

Singularities appear quite generically in classical gravity \cite{PenHaw}. The
cosmic censorship conjecture states that the only singularities of physical
space-times can be: an initial cosmological singularity (big bang), a final
cosmological singularity (big crunch), and singularities resulting from
gravitational collapse, which are hidden behind event horizons \cite{Pen}.
  To be precise, this is just one version of the cosmic censorship conjecture, called
  ``version 2, physical formulation'' in \cite{Wald}, to which we refer for an overview.
  In the case of four-dimensional static space-times, supersymmetry acts as a
  cosmic censor \cite{KalEtAl}, but, since there are non-static stationary
  supersymmetric solutions with naked singularities \cite{GibHulTod}, it is
  clear that supersymmetry alone is not enough to establish cosmic censorship.
  
  A satisfactory quantum theory of gravity should not only be capable of
  establishing cosmic censorship within the semi-classical approximation, but
  also be able to resolve the cosmological and black-hole singularities. String
  and M-theory are currently not able to achieve these goals in general, but
  have already provided a variety of insights into the problem of singularities.
  One new ingredient is the existence of a length scale, $\sqrt{\alpha'}$, at
  which internal string states can be excited. This leads to a very soft UV
  behaviour of scattering amplitudes, which reflects itself in an infinite
  series of higher derivative terms, and in particular higher curvature terms,
  in the low-energy effective action. Although this suggests that curvature
  singularities are smoothed out, it is very hard to make this explicit, except
  in situations where the corresponding space-time can be described by an exact
  conformal field theory.  See for example \cite{GRS} for a recent application
  of such techniques to cosmology.

Besides higher derivative terms, there is another generic
mechanism for avoiding singularities in string and M-theory, 
which one might call
``the intervention of additional states.''
One example of this mechanism are twisted states in toroidal orbifold
compactifications \cite{Orbi}, which prevent the conic singularities of these
spaces to cause singularities of observable quantities.
  Recently, time-dependent orbifolds \cite{HorSte} have become important as models for
  space-like singularities, including cosmological singularities \cite{HorPol,LiuEtAl}.
  See \cite{CorCos} for a review and more references.
  More elaborate versions of the two basic mechanisms take care of the
  geometrical singularities occurring at special points in the moduli spaces of
  Calabi-Yau compactifications.  Such special points are related to flop
  transitions \cite{AGM}, conifold singularities \cite{Str}, conifold
  transitions \cite{GMS} and more general extremal transitions \cite{KMP,KM}.
  Here, in general, both $\alpha'$-corrections and the presence of additional
  light states descending from p-branes wrapped on p-cycles of the internal
  manifold need to be taken into account in order to obtain non-singular
  physical quantities.

While the above examples concern singularities in an internal,
compact manifold, one can also obtain new insights into 
space-time singularities by considering the full string or M-theory
dynamics instead of a naive supergravity approximation. 
A fascinating interplay between internal, compact space and non-compact
space-time is exhibited by the enhan\c{c}on \cite{JPP}. Here, one considers
certain space-time geometries which have a naked curvature singularity in the
supergravity approximation. However, by considering the full string theory one
realises that before the singularity can be reached, particular modes of branes
wrapped on internal cycles become light, and therefore must be taken into
account.  The resulting space-time geometry is then free of naked singularities.
This mechanism, which has been first observed in a specific compactification
with $\Nc$~$=$~$4$ supersymmetry, seems to be quite generic. For instance, a
version of it was discovered in domain-wall solutions of eleven-dimensional
supergravity compactified on a Calabi-Yau three-fold with G-flux
\cite{KalMohShm}.  This compactification describes the bulk dynamics of
five-dimensional heterotic M-theory \cite{Lukas:1998yy, Lukas:1998tt}.
Moreover, it was shown in \cite{Moh} that a similar mechanism prohibits naked
singularities of electric and magnetic BPS solutions of ungauged
five-dimensional supergravity, when embedded into M-theory compactified on a
Calabi-Yau three-fold (without G-flux).  However, except in the case of magnetic
BPS solutions, one has specific examples instead of a general proof.

In this paper we make a step towards a model-independent analysis of space-time
singularities in M-theory compactifications.  The basic idea is to work out
systematically the relation between the geometries of the internal space and of
space-time, and to prove that one always encounters new M-theory physics, such
as additional light states, when, or even before a naked space-time singularity
occurs. Since we work with the dimensionally reduced low-energy effective
action, the geometry of the internal space is encoded in space-time dependent
scalar fields (moduli), while the space-time geometry is obtained by solving the
equations of motion.  We therefore have to relate space-time geometry, in
particular curvature invariants to the geometry of the moduli space. Concretely,
we consider eleven-dimensional supergravity compactified on Calabi-Yau
three-folds and prove that five-dimensional BPS domain-wall solutions (and
electric BPS solutions) cannot have curvature singularities as long as the
moduli fields take values in the interior of the extended K\"ahler cone.  The
only way solutions can become singular is when the boundary of the extended
K\"ahler cone is reached, but there the internal manifold becomes singular and
the description in terms of a five-dimensional effective supergravity theory is
not valid. We will not address the physics of these singularities in this paper,
but make a proposal how they can be approached in the conclusions.  There are
also supergravity solutions with naked singularities which are not related
to boundaries of the extended K\"ahler cone \cite{KalMohShm}.  These are
artifacts of a naive supergravity treatment.  The extended K\"ahler cone is
obtained by gluing together the K\"ahler cones of different Calabi-Yau spaces,
which are related by topological transitions. Along the corresponding internal
boundaries of the extended K\"ahler cone, the effective supergravity lagrangian
is non-singular, but its parameters change through threshold corrections of
M-theory modes which become massless.  If this effect is ignored, one leaves the
M-theory moduli space, and naked space-time singularities can and will occur.
If, however, the effects of the interior boundaries are taken into account
correctly, one moves from the K\"ahler cone of one Calabi-Yau three-fold into
the K\"ahler cone of another one. In this paper we prove that solutions are
non-singular while the moduli take values (i) inside the K\"ahler cone or (ii)
on interior boundaries of the extended K\"ahler cone. Therefore we know that as
long as the scalars are inside the extended K\"ahler cone no naked singularities
occur. If they seem to be present in a naive supergravity treatment, we are
guaranteed to reach an internal boundary of the extended K\"ahler cone before we
reach the singularity.

In other words, the ``enhan\c{c}on-mechanism'' for domain walls and electric BPS
solutions, which was observed in particular models \cite{KalMohShm,Moh}, works
in general. This implies that electric BPS solutions, which are the 
space-times of minimal ADM mass for a given total electric charge,
are always black holes, and never have naked singularities, as long
as the moduli take values inside the extended K\"ahler cone. Similarly,
domain walls can only become singular when the moduli
reach the boundary of the extended K\"ahler cone. 
Thus we establish an, albeit limited, version of cosmic censorship
which might be phrased as ``the K\"ahler cone is a cosmic censor.''  As a
by-product we obtain various nice relations between geometrical quantities of
the K\"ahler cone and of space-time.  We show explicitly how the behaviour of
the metric of the K\"ahler cone on the boundaries is related to the geometrical
degeneration and the new physics occurring there.  Our results confirm that the
interplay between internal and space-time geometry is of central importance in
string theory.  Moreover, when working with effective supergravity actions,
string theory physics is only captured if all relevant modes of the full theory
are taken into account. For more work along these lines see
\cite{Moh,Flop,FlopCos}.

The outline of this paper is as follows.  Section~2.1 introduces domain-wall
solutions of five-dimensional gauged supergravity theories. The analysis of
space-time curvature singularities is performed in section~2.2, and the relation
to Calabi-Yau flux compactifications of M-theory is reviewed in section~2.3.
The main results of this paper reside in section~3, where certain properties of
the Calabi-Yau K\"ahler-cone metric are proven. In section~4 we explain how our
arguments can be applied to electric BPS solutions of ungauged five-dimensional
supergravity. Our conclusions are given in section~5.

\section{BPS Domain-wall Solutions of Five\--\-dimen\-sional Gauged Supergravity}

We consider two classes of 1/2-BPS solutions of five-dimensional $\Nc$~$=$~$2$
supergravity \cite{Gunaydin:1983bi,GST2,Ceresole:2000}, domain-wall, and
black-hole solutions.  Since these solutions have many features in common, we
focus on the domain-wall solutions in the following.  The following arguments
can be adapted to black-hole solutions, as we will show in section~4.

\subsection{Review of Domain-wall Solutions}

In this subsection, we review the domain-wall solution of a class of
five-dimensional gauged supergravity theories, which describes the bulk dynamics
of Ho\v{r}ava-Witten theory compactified on a Calabi-Yau three-fold
\cite{Lukas:1998yy, Lukas:1998tt,Behrndt:2000zh,Greene:2000yb}.

The bosonic fields are part of the following multiplets: Metric and graviphoton,
\{$g_{\mu\nu}$, $A_\mu$\}, belong to the gravity multiplet.  There are $N-1$
vector fields and scalars, \{$A_\mu^i$, $\phi^i$\}, $i$~$\in$~$1\dots N-1$ in
vector multiplets.  Furthermore, we consider the universal hypermultiplet (UHM),
\{$V$, $a$, $\xi$, $\bar{\xi}$\}, consisting of two real and one complex scalar.
The theory might contain additional hypermultiplets which, however, 
do not play a
role in the domain-wall solutions we consider.

The scalar fields $\phi^i$ parametrise a degree-three hyper-surface in
$\mathbbm{R}^N$ \cite{Gunaydin:1983bi, deWit:1991nm}
\begin{equation}
  \label{eq:constraint}
  \Vc(X):=\frac{1}{6}\;c_{IJK}X^IX^JX^K = 1\;,\qquad I,J,K\in{1\dots N}\;,
\end{equation}
determined by the real, symmetric, and constant coefficients $c_{IJK}$.  As the
graviphoton $A$ and the vector multiplet gauge fields $A^i$ combine into $N$
vector fields $A^I$, we combine the $N-1$ scalars $\phi^i$ and $V$ together,
anticipating the structure we will obtain by dimensional reduction, and define
\begin{equation}
  \label{eq:YX}
  Y^I := V^{1/6}X^I\;.
\end{equation}


We consider a particular gauging of the axion in the UHM, which induces a
potential for the moduli. As a consequence, neither flat Minkowski space nor
$AdS_5$ space is a solution. The most symmetric solutions are 1/2-BPS solutions,
invariant under 4 supercharges and under 4-dimensional Lorentz transformations,
only.

The five-dimensional line element of such a domain-wall solution is given by
\cite{Lukas:1998yy, Lukas:1998tt}
\begin{equation}
  \label{eq:DW_metric}
  \fl  
  \rmd s^2 = \exp\big[2U(y)\big]\Big\{
  -(\rmd x^0)^2 +(\rmd x^1)^2+(\rmd x^2)^2+(\rmd x^3)^2 \Big\}  
  +\exp\big[8U(y)\big]\rmd y^2\,,
\end{equation}
in terms of a single function $U$, which only depends on the transversal
coordinate $y$. This function is related to the scalar moduli by
\begin{equation}
  \label{eq:UV}
  \exp\big[6U(y)\big] = V(y) =\left(
    \frac{1}{6}\;c_{IJK}Y^I(y)Y^J(y)Y^K(y)
    \right)^2\;.
\end{equation}
The moduli $Y^I(y)$, in turn, are determined in terms of harmonic functions
$H_I(y)$,
\begin{equation}
  \label{eq:STAB}
  c_{IJK}Y^I(y)Y^K(y) = 2\,H_I(y)\;,\qquad 
  H_I(y)=a_Iy+b_I\;,\quad a_I,b_I\in\Rset\;.
\end{equation}
Note that the domain-wall solution is completely fixed by a flow on the
scalar manifold which is parameterised by the transverse coordinate $y$. The
solution starts at $y$~$=$~$y_1$ at a particular point on the scalar manifold
and evolves as determined by the equations (\ref{eq:DW_metric})--(\ref{eq:STAB})
until it terminates at a different point at $y$~$=$~$y_2$. Since the
five-dimensional theory does not have fully supersymmetric ground states, there
is no fixed-point behaviour and we have to introduce boundaries at the positions
$y_1$, $y_2$ by hand.  The so-called generalised stabilisation equations
\refeq{eq:STAB} are an universal feature of both, domain-wall and black-hole
solutions, and therefore the following analysis of (space-time) curvature
singularities can be adapted for black-hole solutions, as we will show in
section~4.

\subsection{Curvature Singularities of Domain-wall Solutions}

Here, we investigate the occurrence of space-time curvature singularities. We
start by calculating the Ricci scalar of the metric \refeq{eq:DW_metric} and
then analyse possible sources of divergences.  The Ricci scalar is given by
($'=\frac{\rmd}{\rmd y}$)
\begin{equation}
  \label{eq:DW_R}
  R = 4\exp\big[-8U\big]\;\big( 3U'U'-2U'' \big)\;.
\end{equation}
This expression can diverge ($U$ is related to $V$ by equation \refeq{eq:UV}):
\begin{enumerate}
\item if either $\exp\big[-U\big]\rightarrow\infty$ ($V\rightarrow0$),
\item or if the first or second derivatives of $U$ (or $V$) diverge.
\end{enumerate}
Since the line element \refeq{eq:DW_metric} depends only on the function $U$,
all components of the Riemann tensor are polynomials in $U'$ and $U''$.  Hence,
our analysis applies to all curvature invariants of the domain-wall metric.

Since case (i) has already been covered in the literature \cite{Witten:1996qb},
it remains to analyse the somewhat less obvious case (ii), {\it i.e.},~diverging
curvature invariants at finite and non-zero $V$. It is convenient to consider
the first derivative of $\sqrt{V}$ instead of $V$:
\begin{equation}
  \label{eq:V'}
  \Big(\sqrt{V}\Big)' = \frac{1}{2}\;c_{IJK}Y^IY^JY^{\prime K} 
  = \frac{1}{2}\;Y^IH'_I\;,
\end{equation}
where in the last step we have used the relation
\begin{equation}
  \label{eq:STAB'}
  c_{IJK}Y^JY^{\prime K} = H'_I\;,
\end{equation}
which follows from differentiating \refeq{eq:STAB} with respect to $y$.  Since
the harmonic functions $H_I$ are at most linear in $y$, $V'$ is regular as long
as the moduli $Y^I$ are finite \cite{Bergshoeff:2000zn, KalMohShm}.

Differentiating equation \refeq{eq:V'} once more we find
\begin{equation*}
  \Big(\sqrt{V}\Big)'' = \frac{1}{2}\; Y^{\prime I}H'_I + \frac{1}{2}\;Y^IH''_I\;.
\end{equation*}
Clearly, $(\sqrt{V})''$ can blow up if $Y^{\prime I}$ diverges.  By introducing the
matrix \cite{Greene:2000yb}
\begin{equation}
  \label{eq:M}
  \TilM_{IJ} = \frac{1}{2}\;c_{IJK}Y^K\;,
\end{equation}
we can invert equation \refeq{eq:STAB'}:
\begin{equation*}
  Y^{\prime I} = \frac{1}{2}\,\TilM^{IK}H'_K\;.
\end{equation*}
Of course this inversion is only formal, because it requires to compute $\TilM^{IJ}$
as a function of the moduli $Y^I$.
Since $H'_I=a_I=$~const, $|Y^{\prime I}|$~$\rightarrow$~$\infty$ when $\TilM_{IJ}$ is not
invertible, or, equivalently, when $\det\TilM$~$=$~$0$.  Using the last equation,
we obtain
\begin{equation}
  \label{eq:V''}
   \Big(\sqrt{V}\Big)'' = \frac{1}{4}\; H'_I\TilM^{IJ}H'_J + \frac{1}{2}\;Y^IH''_I\;.
\end{equation}
The appearance of the matrix $\TilM_{IJ}$ is the link between space-time curvature
singularities and properties of the moduli-space metric, which we will deal with
in section~3.

We have shown that there are two possible causes for curvature singularities of
domain-wall solutions: (i) $V$~$\rightarrow$~$0$, and (ii) $\TilM_{IJ}$
non-invertible at finite $V$~$\neq$~$0$.  Let us demonstrate that these
curvature singularities do occur \emph{generically}: Since $V$ is a homogeneous
function of degree three in the moduli $Y^I$, it will have zeros if the moduli
$Y^I$ are allowed to take arbitrary real values. This covers case (i).  As for
case (ii), a generic matrix $\TilM_{IJ}$ is invertible, but becomes singular in
co-dimension one in parameter space. In other words, if no additional conditions
on the parameters are imposed, the set of solutions will decompose into two
subsets: those, which do not cross the hyperplanes where $\TilM_{IJ}$ becomes
singular, and those which do.
  This can be seen explicitly 
in the examples considered in Ref.~\cite{KalMohShm}.
In both cases singular and non-singular space-time geometries are equally
generic, and supergravity does not provide any constraints on the parameters
which exclude the singular solutions.
The difference between case (i) and case (ii) is that in 
the first case
the metric on
the scalar manifold diverges, while it develops a zero eigenvalue in the second
case. Thus, the five-dimensional supergravity lagrangian becomes singular, which
indicates that we need input from an underlying fundamental theory.
We will see that this input is provided by M-theory, if the 
supergravity theory is obtained as a Calabi-Yau compactification.


\subsection{Compactification of Eleven-dimensional Supergravity on Calabi-Yau Three-folds with
  Background Flux}

Here, we recall how five-dimensional gauged supergravity 
can be obtained by compactification of eleven-dimensional supergravity on
Calabi-Yau three-folds 
in the presence of background flux.

%
There are two points of view concerning the relation of gauged five-dimensional
supergravity actions to eleven-dimensional supergravity theory: (i)
compactification on a Calabi-Yau manifold, assuming that the flux only excites
Calabi-Yau zero-modes and does not deform the Calabi-Yau structure. The presence
of background flux is taken into account by including it as a ``non-zero mode,''
see Refs.~\cite{Lukas:1998yy, Lukas:1998tt, Behrndt:2000zh}.  Or, (ii),
compactification on a ``deformed'' Calabi-Yau manifold \cite{ Lukas:1998yy,
  Lukas:1998tt, Witten:1996mz, Gurrieri:2002wz}.

We describe the first approach.  The bosonic fields of eleven-dimensional
supergravity consist of the metric and of a three-form gauge potential $C_3$
with associated four-form field strength $G_4=\rmd C_3$.  We start by specifying
a basis of the second cohomology group consisting of $h^{1,1}$ harmonic $(1,1)$
forms $\omega_I$.  The K\"ahler form can be expanded in this basis,
\begin{equation}
  \label{eq:1,1}
  J = \Yc^I\omega_I\;,\qquad \langle\omega_I\rangle = H^{1,1}(X)\;,\quad
  I=1\dots h^{1,1} := \dim H^{1,1}(X)\;,
\end{equation}
with real moduli $\Yc^I$, which are related to the moduli of secion~2 by the rescaling
$\Yc^I$~$:=$~$V^{1/3}X^I$~$=$~$V^{1/6}Y^I$. 
Since we will need a basis of $H^{2,2}(X)$ and of the even
homology of $X$, we introduce dual 4-forms, 2-cycles, and 4-cycles.  By
Poincar\'e duality, there is a dual basis of 4-forms $\nu^I$ defined as
\begin{equation*}
  \int_X \nu^I\wedge \omega_J = \delta^I_J\;,\qquad\langle\nu^I\rangle = H^{2,2}(X)\;.
\end{equation*}
In homology, we fix a basis of 2- and 4-cycles, with relations
\begin{equation*}
  \int_{C^I}\omega_J = \int_{D_J}\nu^I = \delta^I_J
  \;,\quad \langle C^I\rangle = H_2(X)
  \;,\quad \langle D_I\rangle = H_4(X)\;.
\end{equation*}

The symmetric tensor $c_{IJK}$ of section~2 acquires now the interpretation of
triple-intersection numbers
\begin{equation}
  \label{eq:3i}
  c_{IJK} = D_I\circ D_J\circ D_K = \int_X \omega_I\wedge\omega_J\wedge\omega_K\;,
\end{equation}
which implies that it is \emph{integer valued},\footnote{
  This holds in an appropriate basis of (co-)homology and for non-singular $X$. For
  singular Calabi-Yau three-folds these numbers can be rational.
} in contrast to pure five-dimensional supergravity where \emph{real-valued}
tensors are allowed.

Having introduced a basis for the even (co-)homology, we now describe how the
bosonic fields of section~2 descend from the fields of eleven-dimensional
supergravity:
\begin{eqnarray*}
  C_{MNP}\quad&\Longrightarrow\quad A_\mu^I\rmd x^\mu\wedge\omega_I
  \;,\quad \xi\Omega_{abc}
  \;,\quad \bar{\xi}\bar{\Omega}_{\bar{a}\bar{b}\bar{c}}\\
  G_{MNPQ}=(\rmd C)_{MNPQ}\quad&\Longrightarrow\quad
  \rmd a = \star_5\,G\\
  g_{MN}\quad&\Longrightarrow\quad \Yc^I\omega_I
  \;, \quad g_{\mu\nu}
\end{eqnarray*}
Here, $\Omega_{abc}$ denotes the holomorphic $(3,0)$ form which exists on every
Calabi-Yau three-fold. The fields $X^I$ of section~2.1 parameterise the relative
sizes of the cycles of $X$, whereas the UHM scalar $V$ parameterises the volume
of $X$. The axion $a$ comes from dualising the dimensionally reduced 4-form
field strength, and therefore has a shift symmetry: $a\rightarrow a+c$.

In principle, dimensional reduction on a generic Calabi-Yau manifold yields
more hypermultiplets than the UHM alone. For the type of domain-wall solutions we
consider, these extra hypermultiplets are spectators, and it is a consistent
truncation to keep these fields constant.

Following \cite{Lukas:1998yy, Lukas:1998tt, Behrndt:2000zh} we turn on
background flux in a way that the flux back-reaction on geometry is such that it
excites Calabi-Yau zero-modes only, and does not distort the Calabi-Yau
structure.  Since the background four-form flux is an element of
$H^4(X)$~$=$~$H^{2,2}(X)$, it can be expanded as follows
\begin{equation}\label{flux}
  G = \alpha_I\nu^I \in H^{2,2}(X)\;,
\end{equation}
with constants $\alpha_I$ subject to a quantisation condition
\cite{Witten:1996md}.
However, within the supergravity approximation the flux parameters can be taken
to be continuous as discussed in Ref.~\cite{Gurrieri:2002wz}.
In the dimensionally reduced five-dimensional theory, turning on flux
\refeq{flux} leads to (i) a potential for the moduli $\Yc^I$ and (ii) gauging of
the shift symmetry of the axion, $D_\mu a$~$=$~$\partial_\mu
a$~$+$~$\alpha_IA_\mu^I$.
 
It is important to keep in mind that the domain-wall solutions of the last subsections
are exact solutions of five-dimensional gauged supergravity theory, but do not
lift to exact solutions of 
the eleven dimensional theory. 
The corresponding eleven-dimensional domain-wall solutions of Ho\v{r}ava-Witten
theory are only known up to first order, and to this order they agree with the
five-dimensional domain-wall solutions, see~Ref.~\cite{Lukas:1998tt}.

We have already mentioned that the tensor $c_{IJK}$ has to be integer valued in
a Calabi-Yau compactification. Similarly, the scalar fields $\Yc^I$ are subject to
certain constraints we will deal with in the next section.

\section{Properties of the K\"ahler Cone}

Having described five-dimensional domain-wall solutions from the point of view
of supergravity in the last section, we now investigate the interplay between
space-time physics and properties of the K\"ahler moduli space.

\subsection{The K\"ahler Cone of Calabi-Yau Three-folds}

By Wirtinger's theorem, the K\"ahler form measures the volume of holomorphic
curves, surfaces and the volume of the Calabi-Yau manifold $X$. For all
holomorphic curves $C$~$\subset$~$X$ and surfaces $S$~$\subset$~$X$, the
following inequalities define the K\"ahler cone $\Kc$:
\begin{equation}
  \label{eq:KC_constr}
  \eqalign{
  \Vol(C) &= \int_{C} J > 0\;,\\
  \Vol(S) &= \frac{1}{2!}\,\int_{S} J\wedge J > 0\;,\\
  \label{eq:defV}
  V := \Vol(X) &= \frac{1}{3!}\,\int_X J\wedge J\wedge J 
   =\frac{1}{6}\,c_{IJK}\Yc^I\Yc^J\Yc^K = \Vc(\Yc)> 0\;.
  }
\end{equation}
Thus, the K\"ahler moduli space has the structure of a cone.  The (closure of)
the K\"ahler cone is the cone $\overline{N\!E}^1(X)$ of \emph{nef} classes,
which is dual to (the closure of) the Kleiman-Mori cone $\overline{N\!E}_1(X)$
of effective 2-cycles \cite{Wilson, Oda}.
  The duality is given by the pairing $\mathrm{Pic}(X)\times H_2(X)\longrightarrow\Zset$,
  which is $\int_CL$ for a curve $C$ and $L$~$\in$~$\mathrm{Pic}(X)$~$=$~$H^{1,1}\cap
  H^2(X)$, where $\mathrm{Pic}(X)$ denotes the Picard group of $X$.
  If $X$ is a Calabi-Yau three-fold, then the K\"ahler cone is \emph{locally}
  polyhedral away from the so-called cubic cone
  $W$~$:=$~$\big\{\Yc^I\in\Rset\;|\;V=0\big\}$ \cite{Wilson}.  For
  \emph{toric-projective} Calabi-Yau varieties the K\"ahler moduli space is a
  strongly convex finite polyhedral cone \cite{Reid,FJ}, and there is an
  explicit, global parameterisation, which takes the form
\begin{equation}
  \label{eq:KC}
  \Kc := \Big\{\Yc^I\in\Rset\;\Big|\; 0<\Yc^I<\infty\,,\; 1\le I\le h^{1,1}\Big\}\;.
\end{equation}
We call this parameterisation adapted, since the moduli $\Yc^I$ measure volumes of
holomorphic 2-cycles $C_I$.

The metric on the Calabi-Yau K\"ahler moduli space is given by
\cite{Strominger:1990, Bodner:1991}
\begin{equation}
  \label{eq:KC_metric}
  G_{IJ} := \frac{1}{2V}\;\int_X\omega_I\wedge\star\omega_J 
  = -\frac{1}{2}\;
  \frac{\partial}{\partial \Yc^I}\frac{\partial}{\partial \Yc^J}\log \Vc(\Yc)\;.
\end{equation}
This metric is non-degenerate inside the K\"ahler cone.  With the use of
equation \refeq{eq:defV} it can be rewritten as
\begin{equation}\label{eq:GMT}
  G_{IJ} = -\frac{1}{V}\;M_{IK}\left(\delta^K_J - \frac{3}{2}\;T^K_J\right)\;,\qquad
  T^K_J := \frac{1}{6V}\;c_{JMN}\Yc^M\Yc^N\Yc^K\;.
\end{equation}
Here, $M_{IJ}$~$=$~$V^{1/6}\TilM_{IJ}$ is a rescaled version of the matrix $\TilM_{IJ}$
introduced in \refeq{eq:M}.
The matrix $T$ is a
projector, $T^2=T$, of trace one.  By the Hodge index theorem, the signature of
the matrix
\begin{equation*}
  M_{IJ} = \frac{1}{2}\;\int_X J\wedge\omega_I\wedge\omega_J = \frac{1}{2}\;c_{IJK}\Yc^K
\end{equation*}
is $(1,h^{1,1}-1)$ \cite{Wilson}.  Since non-invertability of the matrix $M$ is
one cause of space-time curvature singularities (see~section~2.2), equation
\refeq{eq:GMT} establishes the link between the occurrence of curvature
singularities and properties of the K\"ahler-cone metric.  This connection can
be made more explicit by calculating the determinant of $G$:
\begin{equation}
  \label{eq:detGM}
  \det G 
  = \left(\frac{-1}{V}\right)^{h^{1,1}}\det M \; \det \left(\id-\frac{3}{2}\;T\right)
  = -\frac{1}{2}\left(\frac{-1}{V}\right)^{h^{1,1}}\det M\;,
\end{equation}
where in the last step we have made use of the fact that $T$ is a projector of
trace one.  There is a basis in which $T$ assumes the form
$T$~$=$~diag$(1,0,\dots 0)$, and we obtain
\begin{equation*}
  \det\left(\id - \frac{3}{2}\;T\right) 
  = \det\mathrm{diag}\big(-1/2,1,\dots,1\big) = -1/2\;,
\end{equation*}
which completes the derivation of equation \refeq{eq:detGM}.

It is the aim of the next subsection to use the relation \refeq{eq:detGM} in
order to analyse regularity properties of the metric \refeq{eq:KC_metric}.

\subsection{Degenerations of the K\"ahler-cone Metric and Singularities
  of Space-time} In this subsection, we analyse how the K\"ahler-cone metric
\refeq{eq:KC_metric} behaves on boundaries of the K\"ahler cone, in particular
whether it develops zero eigenvalues.
  By ``K\"ahler-cone metric at the boundary'' we always mean the limit of the
  K\"ahler-cone metric as one approaches the boundary, and not the scalar metric of the
  extended effective field theories which explicitly include the additional light modes
  \cite{Witten:1996qb,Flop,FlopCos}.
  
We consider boundaries of the K\"ahler cone where one particular 2-cycle,
$C^\star$, collapses:
\begin{equation}
  \label{eq:KC_boundary}
  \partial_\star\Kc := \left\{\left(\Yc^\TilI\neq0,\; \Yc^\star=0\right)\,,\quad
    0<\Vc(\Yc)<\infty\right\}\;,\quad \TilI\neq\star\;.
\end{equation}
The contractions at these co-dimension-one faces are called primitive.  In
Calabi-Yau three-folds the following contractions can take place \cite{Wilson,
  Witten:1996qb}:
\begin{itemize}
\item Type I (``$2\rightarrow0$''): A finite number of isolated curves in the homology
  class $C^\star$ is blown down to a set of points, $\Vol(C^\star)=\Yc^\star\rightarrow0$,
  {\it e.g.}, (locally) $\Oc(-1)\oplus\Oc(-1)\longrightarrow\mathbbm{P}^1$.
\item Type II (``$4\rightarrow0$''): A divisor $D=v^ID_I$ collapses to a set of points:
  $\Vol(D)$~$\propto$~$(\Yc^\star)^2$.
\item Type III (``$4\rightarrow2$''): A (complex) one-dimensional family of curves sweeps
  out a divisor $D=v^ID_I$. Contracting this family of curves induces a collapse
  of $D$ to a curve of genus $g$: $\Vol(D)$~$\propto$~$\Yc^\star$, {\it e.g.},
  ($g=0$ case) $\Oc(0)\oplus\Oc(-2)\longrightarrow\mathbbm{P}^1$.  
\item Cubic cone (``$6\rightarrow4$'', ``$6\rightarrow2$'', ``$6\rightarrow0$''): These
  contractions correspond to $V$~$\propto$~$\Yc^\star$,
  $V$~$\propto$~$(\Yc^\star)^2$ and $V$~$\propto$~$(\Yc^\star)^3$.
\end{itemize}
Note that our definition of boundaries $\partial_\star\Kc$ in equation
\refeq{eq:KC_boundary} does not include the last of these possibilities.  This
is already covered by case (i) in section~2.2: the Ricci scalar
($R$~$\propto$~$V^{-8/3}$) diverges when $V\rightarrow0$.

Boundaries of type I and type III can be crossed into the
K\"ahler cone of a new Calabi-Yau
threefold, which is birationally (and, for type III, even biholomorphically)
equivalent to the original one.  
Crossing these boundaries corresponds to a flop \cite{AGM} or going through 
gauge symmetry enhancement
\cite{Wilson, KMP}, respectively.
The extended K\"ahler
cone is gotten by enlarging the K\"ahler moduli space at all boundaries of type I.
Enlarging in addition the K\"ahler moduli space at all boundaries of type III,
one obtains
the extended movable cone \cite{AGM, KMP}. However, this second extension 
only adds ``gauge copies'' to the
parameter space (see for example \cite{Flop,Moh} for an explanation).
While type-I and type-III boundaries are ``internal boundaries'' 
of the M-theory
moduli space, type-II contractions and the cubic cone 
lead to proper boundaries.
At boundaries of type II
the M-theory moduli
space ends, and it has been shown 
that the tension of strings descending from M5-branes wrapped on the divisor 
goes to zero at such boundaries\cite{Witten:1996qb}. Here the
supergravity approximation breaks down, because infinitely many M-theory states
become massless. Similarly, the supergravity approximation breaks down when
approaching the cubic cone, and in this case no interpretation in terms of
M-theory physics is known.\footnote{
  However, when dimensionally reducing on the M-theory cycle, such regions correspond to
  non-geometrical phases of type-IIA string theory on the same Calabi-Yau manifold
  \cite{Witten:1996qb,Witten:1993}.
}

Using equation \refeq{eq:detGM}, for finite and non-zero Calabi-Yau volume $V$,
we are able to infer regularity properties of the K\"ahler-cone metric $G$ from
the matrix $M$ and vice versa: there is a one-to-one map of zero eigenvalues of
$G$ to zero eigenvalues of $M$, {\it i.e.},~if
\begin{equation*}
  \det(M_{IJ})\big|_{\Yc^\star\rightarrow0}\propto(\Yc^\star)^n\;,
\end{equation*}
then there are $n$ linearly independent eigenvectors of $M$ (and of $G$)
satisfying
\begin{equation}\label{eq:zeroEVs}
  v^I_{(i)}M_{IJ}\big|_{\Yc^\star=0} = 0\;,\quad i=1\dots n\;.
\end{equation}
Here and in the following, $|_{\Yc^\star\rightarrow0}$ denotes the limit
approaching the boundary $\partial_\star\Kc$.
%
Equation \refeq{eq:zeroEVs} is supposed to hold throughout the face
$\partial_\star\mathcal{K}$.  
In particular, the null
eigenvectors are determined by the triple intersection numbers, only.
This implies that the components of the
eigenvectors can be chosen to be \emph{integer}. 
%
%
Hence, each zero eigenvector $v_{(i)}^I$ defines a divisor
\begin{equation*}
  D_{(i)} := v^I_{(i)}D_I\;.
\end{equation*}
If there is a holomorphic surface within the homology class $D_{(i)}$, then its
volume is given by
\begin{equation*}
\fl  \eqalign{
     \frac{1}{2}v_{(i)}^I\int_X\omega_I\wedge J\wedge J&= 
    \frac{1}{2}\,v_{(i)}^I\,c_{I\TilJ\TilK}\,\Yc^\TilJ \Yc^\TilK
    + v_{(i)}^I\,c_{I\TilJ\star}\,\Yc^\TilJ \Yc^\star
    + \frac{1}{2}\,v_{(i)}^I\,c_{I\star\star}\,\Yc^\star \Yc^\star\\
    & = 
    v_{(i)}^IM_{I\TilJ}\big|_{\Yc^\star=0}\,\Yc^\TilJ 
    + 2\,v_{(i)}^IM_{I\star}\big|_{\Yc^\star=0}\,\Yc^\star
    + \frac{1}{2}\,v_{(i)}^I\,c_{I\star\star}\,\Yc^\star \Yc^\star\\
    &= \frac{1}{2}\,v_{(i)}^I\,c_{I\star\star}\,\Yc^\star \Yc^\star\;,
  }
\end{equation*}
where we have used equation \refeq{eq:zeroEVs}.  As a consequence, the divisors
$D_{(i)}$,
which are associated to null eigenvectors $v_{(i)}$,
can never perform a type-III contraction, which is characterized by
$\Vol(D)$~$\propto$~$\Yc^*$.
Irrespective of whether there exists a holomorphic surface in the class
$D_{(i)}$, we learn that the moduli-space metric is always regular at boundaries
of type I and type III.  On the other hand, by definition of a 
type-II boundary
(``$4\rightarrow0$'', {\it i.e.~}$\Vol(D)$~$\propto$~$(\Yc^\star)^2$),
we know that there exists at least one surface with
homology class $D=v^ID_I$, which collapses to a point at $\partial_\star\Kc$. 
Hence the moduli-space metric
develops a zero eigenvalue at boundaries of type II.  At the cubic cone, the
determinant of the moduli space generically diverges.
  More precisely, there are two cases: If $V$~$\propto$~$(\Yc^\star)^3$, or
  $V$~$\propto$~$(\Yc^\star)^2$, then the determinant of $G$ always diverges,
  see~equation~\refeq{eq:detGM}, since the determinant of the matrix $M$ can never compensate
  the zero in the denominator. What happens in the remaining case, $V$~$\propto$~$\Yc^\star$
  is that \emph{generically} the determinant of $G$ blows up, while at special points
  $\det M$ can compensate the zero in the denominator of equation \refeq{eq:detGM}.
  Table~\ref{BoundRes} summarizes our result.
\begin{table}
  \caption{Behaviour of K\"ahler moduli-space metric at boundaries of the K\"ahler
    cone. 
  }
  \begin{indented}
  \item[] \begin{tabular}{cc}
      \br
      type of boundary & behaviour of $\det(G_{IJ})$ \\
      \mr
      type I  & regular\\
      type II & zero\\
      type III & regular\\
      cubic cone & divergent\\
      \br
    \end{tabular}
    \label{BoundRes}
  \end{indented}
\end{table}

The generalisation of the proof to non-toric Calabi-Yau manifolds, where the
global parameterisation of the K\"ahler moduli space which we have used above need
not exist, is as follows.  As in the proof above, at the cubic cone $W$ the
volume of $X$ vanishes ({\it i.e.}, $V$~$=$~$0$) and generically curvature
singularities occur in domain wall solutions. Moreover, the K\"ahler cone
metric also diverges at the cubic cone.  Since the
K\"ahler cone is locally polyhedral away from $W$, we know that for each of the
primitive faces there exists a local parameterisation of the form
\refeq{eq:KC_boundary}. There can be accumulation points of faces, but these are
known to reside inside the cubic cone \cite{Wilson}.  Thus, the proof is valid
for \emph{all} Calabi-Yau three-folds.

Now we are able to interpret the singularities of section~2 in terms of M-theory
physics. Singularities of type (i) correspond to the cubic cone where the volume
of the Calabi-Yau three-fold goes to zero.  Singularities of type (ii) can occur
in two different situations: The first is that one has reached a type-II
boundary ({\it i.e.}, $\det G$~$=$~$0$).  On these boundaries the internal
manifold and the effective supergravity lagrangian become singular, and
tensionless strings appear, as discussed above.  However, there is also the
possibility that a singularity of type (ii) arises because one has crossed a
boundary of type I or type III before, so that one is outside the K\"ahler cone.
This situation is analogous to the enhan\c{c}on mechanism \cite{JPP}. When
reaching boundaries of type I or type III, the triple-intersection numbers and
therefore the low-energy equations of motion and the space-time metric change.
Continuation of domain-wall solutions through type-I boundaries have been
considered in Ref.~\cite{Greene:2000yb}, whereas continuation of black-hole and
black-string solutions through type-I and type-III boundaries have been studied
in Ref.~\cite{Moh}.  Here we only need to use that type-I and type-III
boundaries are internal boundaries of the extended K\"ahler cone, and that the
metric of the extended K\"ahler cone does not become singular.  After crossing
such boundaries the moduli take values in another K\"ahler cone, and there our
proof of absence of naked singularities applies again. In conclusion we see that
in M-theory singularities only occur on the boundary of the extended K\"ahler
cone, where the internal manifold and the five-dimensional effective lagrangian
become singular, and the description in terms of five-dimensional supergravity
breaks down.

\subsection{Example: The $\mathcal{F}_1$-Model}

\begin{figure*}[t]
    \centerline{\epsfbox{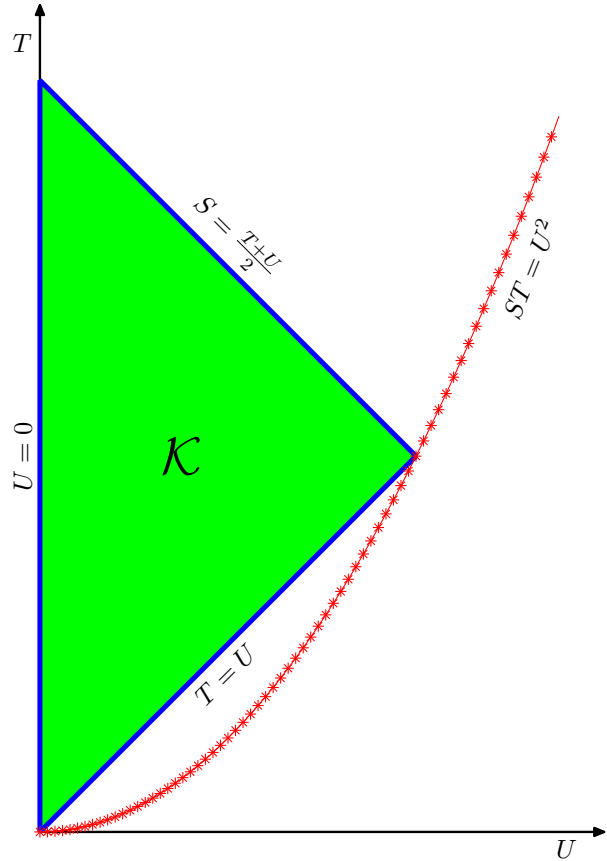}}
  \caption{\label{fig:KC} Section of the $\mathcal{F}_1$-model K\"ahler cone for fixed
    modulus $S$~$>$~$0$.} 
\end{figure*}

Here, we present a well known example of a Calabi-Yau manifold with
$h^{1,1}$~$=$~$3$ \cite{Louis:1996mt,ChouEtAl}, which has all features discussed
in the last subsections. It is an elliptic fibration over the first Hirzebruch
surface ${\cal F}_1$.  It turns out to be convenient to choose the following
non-adapted parametrisation of the K\"ahler cone:\footnote{
  In this subsection, $U$ denotes one of the scalar fields, and not the function appearing
  in the space-time metric.
}
\begin{equation*}
\Kc = \left\{ 
    S,T,U\in\Rset \;\Big|\; T>U>0\;,\quad S>{{T+U}\over2}
  \right\}
  \;,\quad 
  V = STU + \frac{1}{3}\,U^3\;.
\end{equation*}
The matrices $M_{IJ}$ and $G_{IJ}$ take the form
\begin{equation*}
\fl  M = \frac{1}{2}\;\left(
  \begin{array}{ccc}
    2U & S & T\\
    S & 0 & U \\
    T & U & 0
  \end{array}\right)\;,
  \qquad
  G= \frac{1}{6V^2} \;
  \left(
  \begin{array}{ccc}
    U^4+3T^2S^2 & 2SU^3 & 2TU^3\\
     2SU^3 & 3U^2S^2 & -U^4\\
     2TU^3 & -U^4 & 3U^2T^2
   \end{array}
   \right)
 \end{equation*}
with determinants 
\begin{equation*}
  \det M = \frac{U}{4}\,\big(ST-U^2)\;,\qquad
  \det G = \frac{\det M}{2V^3}\;,
\end{equation*}
satisfying equation \refeq{eq:detGM}.  Figure~\ref{fig:KC} displays the K\"ahler
cone of this model and table~\ref{tab} summarises the information in
figure~\ref{fig:KC} \cite{ChouEtAl,Flop,Moh}.  Note that the curves with $\det
G$~$=$~$\infty$ ($U$~$=$~$0$) and with $\det G$~$=$~$0$ ($ST$~$=$~$U^2$), lie
always outside or at boundaries of $\Kc$, in accord with the general statement
of table~\ref{BoundRes}. The line $ST$~$=$~$U^2$ is called discriminant line,
because $ST$~$-$~$U^2$ is the discriminant of a specific polynomial, whose zeros
are in one-to-one corresponce with diverging derivatives of the scalar fields
\cite{Moh}. As long as $ST$~$>$~$U^2$, this polynomial does not have real zeros,
and derivatives of scalar fields cannot diverge. Observe that the discriminant
line lies beyond the type-III boundary, where gauge symmetry is enhanced.
Therefore the naked space-time singularities occurring at $ST$~$=$~$U^2$ are
unphysical \cite{KalMohShm,Moh}.
  For the analogous black hole solution,
  the correct non-singular continuation beyond
  the type-III boundary is described in \cite{Moh}.
  
The extended K\"ahler cone of the model is obtained by extending it along the
flop line, $S$~$=$~$(T+U)/2$.  The flopped image of $\Kc$ has boundaries of
type II, where
the metric degenerates. \\
\begin{table}
    \caption{$\mathcal{F}_1$-model K\"ahler cone, {\it cf.}~figure \ref{fig:KC} }
  \begin{indented}
    \item[]\begin{tabular}{@{}r@{}llll}
      \br
      & location & name & physics & $\det (G_{IJ})$\\
      \mr
      $U$&~=~$0$              & cubic cone &unknown &divergent\\
      $T$&~=~$U$              & type I & flop transition &regular\\
      $S$&~=~$\frac{T+U}{2}$  & type III & $SU(2)$ symmetry enhancement &regular \\
      $ST$&~=~$U^2$           & discriminant line & unphysical &degenerate\\
      \br
    \end{tabular} 
    \label{tab}
  \end{indented}
\end{table}

\section{Electric BPS Solutions of Ungauged Five-dimensional
Supergravity}

Let us now indicate how our analysis can be adapted to electric BPS solutions,
which, when no naked space-time singularity occurs, are black holes. Here we
consider a different five-dimensional supergravity theory, namely ungauged
supergravity with an arbitrary number of vector and hypermultiplets, which
admits five-dimensional Minkowski space as a fully supersymmetric solution.
Since hypermultiplets are trivial in electric BPS solutions, we can ignore them
and use the general vector multiplet lagrangian of \cite{Gunaydin:1983bi}.
Again this sector is completely determined by a real cubic prepotential.
Five-dimensional ungauged supergravity with vector and hypermultiplets can be
obtained by dimensional reduction of M-theory on a Calabi-Yau three-fold without
flux \cite{CCDF,Ferrara:1996hh}.  As before, the vector multiplet scalars
parametrise a cubic hyper-surface in the K\"ahler cone, and the coefficients
$c_{IJK}$ are the triple-intersection numbers.

Electric BPS solutions of ungauged five-dimensional supergravity have been found
in \cite{CS} (who use conventions slightly different from those of
\cite{Gunaydin:1983bi}).  The line element is
\begin{equation*}
  \rmd s^2 = - \exp\left[-4 U(r)\right]\rmd t^2 
  + \exp\big[2 U(r)\big] \Big\{\rmd r^2 + r^2 \rmd\Omega_{(3)}^2 \Big\} \;,
\end{equation*}
and the non-vanishing components of the field strengths take the form
\begin{equation*}
  F^I_{tr} = - \der_r\Big( \exp\big[-3 U(r)\big]\, Y^I(r)\Big) \;.
\end{equation*}
Here $Y^I(r)$~$=$~$\exp\big[U(r)\big]X^I(r)$~$=$~$\exp\big[U(r)\big]V^{-1/3}\Yc^I$ 
are again rescaled scalar fields,
subject to the condition ${\cal V}(Y) = \exp\big[3 U(r)\big]$. The full set of
equations of motion and Killing spinor equations reduces to the same set of
algebraic equations as for domain walls ({\it c.f.~}(\ref{eq:STAB})):
\begin{equation}
c_{IJK}\, Y^J(r)\, Y^K(r) = 2\,H_I(r) \;,
\label{StabBH}
\end{equation}
where this time the functions $H_I(r)$ are harmonic with respect to
four-dimensional space. For the above line element, which describes a
single-centered solution, these take the form
\begin{equation*} 
  H_I (r) = c_I + \frac{q_I}{r^2} \;,
\end{equation*}
where $r$ is the radial coordinate, $q_I$ are the electric charges carried by
the solution, and $c_I$ determine the values of the moduli at infinity. It is
remarkable that two \emph{different space-time geometries}, which arise as
solutions of two \emph{different supergravities}, are governed by the \emph{same
  dynamical system} on their vector multiplet manifolds (which agree, if
$c_{IJK}$ are the same). This becomes more transparent through dimensional
reduction to $0+1$ dimensions \cite{BGS}.

The Ricci scalar corresponding to the above line element takes the form
\begin{equation*}
R = - 2r^{-1}\exp\big[-2U\big]\Big( rU''+3r(U')^2+3U'\Big)\;.
\end{equation*}
In the coordinates we have chosen, the solution becomes asymptotically flat for
$r$~$\rightarrow$~$\infty$, whereas $r=0$ is either the event horizon of a
black hole, or a singularity. The limit $r$~$\rightarrow$~$0$ is related to the
celebrated black-hole attractor mechanism, which was first discovered in
four-dimensional extended supersymmetry \cite{BHattractor}, but also occurs in
five-dimensional supergravity \cite{CS}. BPS solutions of supergravities with
eight supercharges have four Killing spinors, but if one imposes that they
behave regularly at $r$~$=$~$0$, then the number of Killing spinors must double
in the limit $r$~$\rightarrow$~$0$.  This implies that the solutions
interpolate between two fully supersymmetric solutions, flat space at
$r$~$\rightarrow$~$\infty$, and $AdS^2$~$\times$~$S^2$ or
$AdS^2$~$\times$~$S^3$ at $r$~$\rightarrow$~$0$, depending on whether the
total number of space-time dimensions is four or five.  These geometries
describe the event horizons of black holes, which are generalisations of the
extreme Reissner-Nordstr\"om solution. Moreover, the scalar fields take discrete
fixed point values at the horizon, which are determined by the so-called
stabilisation equations.  The attractor mechanism is crucial for understanding
black-hole entropy along the lines of \cite{StrVaf}, because it implies that the
Bekenstein-Hawking entropy does not change when varying the values of scalar
fields at infinity. While the original work of \cite{BHattractor} focused on the
behaviour of four-dimensional black holes close to their horizon, it soon became
clear that the corresponding full black-hole solutions are determined by a
rescaled version of the stabilisation equations, which have been called
generalised stabilisation equations \cite{BH}. Equation (\ref{StabBH}) is the
five-dimensional version of the generalised stabilisation equation, and the
stabilisation equation determining the geometry at $r$~$\rightarrow$~$0$ can
be found by a suitable scaling limit \cite{CS}.
  In four dimensions it has been shown that the generalised stabilisation 
equations are not
  only sufficient, but also necessary for having a supersymmetric solution \cite{CDKM}.
  It was also shown that one cannot switch on non-trivial hypermultiplets, but that it is
  possible to include the effect of certain higher curvature terms \cite{CDKM}.  We refer
  to \cite{MohHabil} for a review.  These results should survive in the
  de-compactification limit and therefore also apply to five-dimensional black holes.

Note that the supersymmetric attractor mechanism, which we just reviewed, does
not occur in the domain-wall solutions of Ho\v{r}ava-Witten theory. The reason is
that the five-dimensional bulk supergravity theory does not have fully
supersymmetric solutions, due to the runaway of the volume modulus
\cite{Lukas:1998yy,Lukas:1998tt,Behrndt:2000zh}.  Therefore these domain walls
do not interpolate between vacua, but have to be cut by introducing boundaries.
Even though there is no fixed-point behaviour in these solutions, they are
nevertheless determined by the generalised stabilisation equations
\refeq{eq:STAB}.  There are other five-dimensional gauged supergravities, in
particular those obtained by gauging the R-symmetries of vector multiplets
\cite{GST2}, which have domain walls interpolating different AdS vacua.  For
those the attractor mechanism applies, as pointed out in
\cite{Bergshoeff:2000zn}. However, it is not known how to obtain these
supergravities as compactifications of string or M-theory.  One interesting
speculation is that the run-away behaviour of the volume in Ho\v{r}ava-Witten domain
walls stops, once a minimal volume of the Calabi-Yau space has been reached
\cite{Behrndt:2000zh}. This has found some support recently by new results about
loop corrections to the universal hypermultiplet \cite{AMTV}.

Let us now consider the question of naked singularities in 
electric BPS solutions. Since the behaviour at $r$~$=$~$0$ is
taken care of by the attractor mechanism, we only 
need to analyse under which conditions a solution can develop 
a naked singularity for finite
values $r$~$=$~$r_S$, $\infty$~$>$~$r_S$~$>$~$0$ of the radial variable. 
As for domain walls, the solution is determined by a flow on the
vector-multiplet manifold, which is now parametrised by $r$,
$\infty$~$>$~$r$~$>$~$0$, instead of $y$. Whereas the parameters $c_I$ determine
the behaviour for $r$~$\rightarrow$~$\infty$, the parameters $q_I$ determine
it for $r$~$\rightarrow$~$0$.  By inspection of the Ricci scalar, and of
other curvature invariants, a curvature singularity occurs if and only if either
$U$~$=$~$0$, or if $U'$ or $U''$ diverge at $r$~$=$~$r_S$.  Since $r$~$>$~$0$,
the harmonic functions $H_I$ and all their derivatives are finite, and the
analysis done for domain walls goes through. Thus we find again that 
in supergravity regular and singular solutions are equally generic,
while solutions in Calabi-Yau compactifications cannot become singular, 
as long as the scalar fields take values inside the extended K\"ahler cone. 
Singularities do occur when the moduli reach the boundary of the extended 
K\"ahler cone.

For completeness, let us finally mention further details which are different
from the case of domain walls, but fortunately do not interfere with our
analysis.  These differences occur because in ungauged supergravity vector and
hypermultiplets only couple through gravity, while there are gauge couplings and
a scalar potential in gauged supergravity.  As we have already mentioned the
hypermultiplets are trivial in electric BPS solutions. Since the overall volume
of the internal space sits in a hypermultiplet, this implies that it is
constant, in contrast to the domain-wall case, where it is a specific function
of the vector-multiplet moduli \refeq{eq:UV}. This has some impact on the
discussion of the cubic cone, where the Calabi-Yau volume goes to zero. While
this locus can be approached in domain-wall solutions, it cannot in black-hole
solutions, because the volume is fixed.  Nevertheless, there is a related
degeneration occurring in the black-hole case, where some cycles go to zero
while other go to infinity, in such a way that the total volume is constant.
This happens for example at one boundary of the K\"ahler cone of the ${\cal
  F}_1$-model \cite{Moh}. Another minor difference is that, as we have seen in
section~2, for domain walls one can prove that $U'$ cannot diverge if $U$ is
finite.  The difference is that derivatives of the linear functions $H_I(y)$ are
constant or vanishing, while the derivatives of the functions $H_I(r)$ are not
constant, though still finite. As a consequence, in the black-hole case $U'$ can
become singular for finite $U$, but this can only happen outside the extended
K\"ahler cone, or at points on its boundary.

\section{Conclusions}

We have analysed domain-wall solutions \refeq{eq:DW_metric} of five-dimensional
gauged supergravity theories. Our analysis applies, with only minor
modifications, to black-hole solutions of ungauged supergravity.  In particular,
we have investigated the appearance of curvature singularities in
domain-wall solutions and found that there are two possible causes, namely
$V$~$\rightarrow$~$0$ and $|V''|$~$\rightarrow$~$\infty$.  Within a pure
supergravity perspective this is all what can be achieved.

Embedding these theories into a higher-dimensional theory, {\it i.e.},
eleven-dimensional supergravity on a Calabi-Yau flux background, changes the
situation: the five-dimensional scalar fields become identified with volumes of
2-cycles in the Calabi-Yau three-fold. Therefore, they are required to take
values inside the K\"ahler cone of the Calabi-Yau manifold.

We have proven that naked curvature singularities 
cannot occur as long as the scalars take values inside the 
extended K\"ahler cone. Singularities do appear when the scalars
reach the boundary, where the internal manifold and the 
five-dimensional effective 
supergravity action become singular, and where one needs to 
understand new M-theory physics, such as tensionless
strings.

Furthermore, we have obtained model-independent information on the behaviour of
the K\"ahler moduli-space metric near boundaries of the K\"ahler cone.  We have
proven that the moduli-space metric is regular at boundaries of type I and of
type III, whereas it develops zero eigenvalues at boundaries of type II and
diverges at the cubic cone, see~table~\ref{BoundRes}.

Our result is consistent with the fact that one can extend the K\"ahler cone at
boundaries with type-I and type-III contractions by gluing the K\"ahler cone of
a different Calabi-Yau manifold to the corresponding face
\cite{Witten:1996qb,KMP,Flop}.  Moreover, it complements the other important
model-independent result about black holes in Calabi-Yau compactifications of
M-theory, which states that attractor points are unique in the extended K\"ahler
cone \cite{Wijnholt:1999vk}.

Results about other geometries support the idea that the enhan\c{c}on-like
mechanism, established in this paper for domain walls and electric BPS
solutions, is much more generic.  In particular, ungauged five-dimensional
supergravity also has magnetic BPS solutions \cite{CS}. For these it is very
simple to show that they cannot have naked singularities as long as the moduli
take values in the extended K\"ahler cone \cite{Moh}. Moreover, it was observed
that cosmological solutions of Kasner type 
do not become singular, as long as the moduli take values in the extended 
K\"ahler cone
\cite{FlopCos}.

A natural next step is trying to establish a similar mechanism for black holes
of four-dimensional $\Nc$~$=$~$2$ supergravity.  Since in this case one has to
deal with a holomorphic instead of a real cubic prepotential, this will require
a non-trivial extension of the framework used in this paper. When considering
type-IIA string theory on a Calabi-Yau three-fold, which is the dimensional
reduction of the M-theory setup employed in this paper, then the moduli space
is a complexified version of the K\"ahler cone.  Now the overall volume $V$ sits
in a vector multiplet, and the metric on the K\"ahler cone gets
$\alpha'$-corrections. Moreover, the former boundaries of the K\"ahler cone now
become so-called non-geometric phases, which are not described by Calabi-Yau
sigma models, but by other types of world-sheet conformal field theories
\cite{Witten:1993,Witten:1996qb}.  This is a technical complication, but a
physical bonus, because now the cubic cone is on the same footing as the other
degenerations.  For example, the K\"ahler cone of the ${\cal F}_1$-model has a
boundary which belongs to the cubic cone \cite{Moh}. After dimensional
reduction, this region corresponds to a Landau-Ginsburg theory.\footnote{
  We thank Albrecht Klemm for pointing this out.
} Moreover, one can use mirror symmetry to study IIB string theory on the mirror
Calabi-Yau manifold instead. In this description the relevant scalar manifold is
the moduli space of complex structures, which does neither get $\alpha'$ nor
string loop corrections. Thus the best strategy is to investigate the links
between complex-structure moduli spaces of Calabi-Yau spaces and BPS space-time
geometries. This is a vast and very interesting subject, and some of its aspects
have already been explored \cite{Moore:1998,Denef:1998sv}. Singularities of the
complex moduli space occur in complex co-dimension one, only. Therefore they will
be avoided by generic solutions. Many of the singularities in co-dimension one
and higher have known physical interpretations in terms of additional massless
states coming from wrapped D-branes.  Since D-branes can be treated by
world-sheet techniques, one now has additional tools besides effective
supergravity. In particular, the best strategy to understand the M-theory
physics of the cubic cone and of tensionless strings is the dimensional
reduction to type II string theory.

A point which deserves further investigation is the relation of supergravity
actions to string compactification on singular Calabi-Yau manifolds.  Two
different questions can be investigated in this context: First, we would like
to know what are the supergravity lagrangians corresponding to the
``$4\rightarrow0$'' boundary theories, {\it i.e.},~which supergravity
lagrangians correspond to the non-geometric phases.  And secondly, it would
be interesting to use the supergravity description in order to learn about the
cohomology and intersection homology of the singular Calabi-Yau three-folds
obtained by type I, II and III contractions.

\ack
This work is supported by the `Schwerpunktprogramm Stringtheorie' of the DFG.

\end{document}